\documentclass[11pt]{article}
\usepackage{amsmath}

\numberwithin{equation}{section}

\author{
  Christian G.~B\"ohmer\thanks{c.boehmer@ucl.ac.uk}\, and 
  Robert J.~Downes\thanks{r.downes@ucl.ac.uk}\\ \\
  ${}^{\ast}$Department of Mathematics, \\ 
  ${}^{\dagger}$Centre for Advanced Spatial Analysis, \\
  University College London, \\
  Gower Street, London, WC1E 6BT, UK
}

\title{From continuum mechanics to general relativity}
\date{19 May 2014}

\begin{document}

\maketitle

\begin{center}
This essay received a honorable mention in the 2014 essay competition of the Gravity Research Foundation.
\end{center}
\mbox{}\\\mbox{}

\begin{abstract}
Using ideas from continuum mechanics we construct a theory of gravity.  We show that this theory is equivalent to Einstein's theory of general relativity; it is also a much faster way of reaching general relativity than the conventional route.  Our approach is simple and natural: we form a very general model and then apply two physical assumptions supported by experimental evidence.  This easily reduces our construction to a model equivalent to general relativity.  Finally, we suggest a simple way of modifying our theory to investigate non-standard space-time symmetries.
\end{abstract}

\vfill

\clearpage

\begin{quote}
    \emph{Einstein himself, of course, arrived at the same Lagrangian but without the help of a developed field theory, and I must admit that I have no idea how he guessed the final result.  We have had troubles enough arriving at the theory - but I feel as though he had done it while swimming underwater, blindfolded, and with his hands tied behind his back!}

    \hfill Richard P.~Feynman, \emph{Feynman lectures on gravitation}, 1995.
\end{quote}

\section{Introduction}

Einstein's theory of general relativity is one of the crowning achievements of modern science.  However, the route from a special-relativistic mind game to a fully-fledged theory capable of producing physical predictions is long and tortuous. As Feynman notes, Einstein completed this journey with seemingly superhuman intuition: for the rest of us, lengthy mathematical preparation is the key to developing the theory.

The purpose of our essay is to showcase a new approach to general relativity.  Using well-known machinery from continuum mechanics~\cite{JaunzemisLove} we construct a general Lagrangian which describes a space-time continuum.  We make two physical assumptions which reduce our Lagrangian (relatively painlessly) to that of general relativity, the Einstein-Hilbert action.  We also suggest a simple way of modifying our theory to investigate different models of gravitation with alternative physical assumptions.

Readers with a background in continuum mechanics should be familiar with our construction while this essay serves as an introduction to general relativity for members of this community.

\section{Mathematical preliminaries}

There is strong experimental evidence that space and time should not be treated as independent concepts, but that we should think of a $4$-dimensional space-time continuum when describing nature.  Such a continuum is modeled mathematically as a $4$-dimensional manifold equipped with Lorentzian metric
\begin{equation}
\label{1:1}
ds^2
=
\eta_{ij}\delta^i_\alpha\delta^j_\beta dx^\alpha dx^\beta
:=
(dt)^2
-
(dx^1)^2
-
(dx^2)^2
-
(dx^3)^2
\end{equation}
which is used to determine distances between points. We adopt Einstein's summation convention whereby summation is carried out over repeated indices. Here $x^\kappa$, $\kappa=1,2,3$ are the standard Cartesian coordinates and $x^0=t$ is time. This is the usual setting for special relativity, a theory containing no gravity. The Kronecker $\delta$'s (where $i,j=0,1,2,3$) indicate that all local coordinate frames are aligned. This over-complication will become useful to us further on.

Light propagating in a vacuum described by space-time (\ref{1:1}) travels in straight lines.  However, experimental evidence suggests that in our universe light rays are deflected by massive bodies.  We argue that this observation indicates (\ref{1:1}) should be modified when gravity is taken into account.  To do this we assume that our local coordinate system changes from point-to-point on the manifold in the presence of gravity.  Mathematically, we modify our metric by attaching to every point a \emph{frame} or \emph{tetrad} $e^i{}_\alpha(x^\kappa,t)$, which describes this change of coordinates
\begin{equation}
\label{1:2}
ds^2
=
\eta_{ij}e^i{}_\alpha e^j{}_\beta dx^\alpha dx^\beta
:=
g_{\alpha\beta}dx^\alpha dx^\beta
.
\end{equation}
The sixteen components of the frame can be thought of as a set of orthonormal vectors attached to each point of the manifold, where $i$ labels each vector and $\alpha$ their components.  At any fixed point $P$ of our manifold, the functions $e^i{}_\alpha$ determine how the local coordinates at $P$ are related to the local coordinates at any other point $Q$.

The frame will play the role of the dynamical variable.  We are therefore modelling a situation in which physical information related to the gravitational field is encoded in the frame.  Gravity manifests itself as the difference in alignment of local coordinates at different points; this is the reason we used Kronecker $\delta$'s in (\ref{1:1}) when dealing with the special-relativistic case.

We can also view the frame as a pseudo-orthogonal matrix.  This suggests that we can introduce the inverse frame $e_j{}^\beta$ which, together with the frame itself, will satisfy the equalities
\begin{equation}
\label{1:3}
e^i{}_\alpha e_i{}^\beta=\delta_\alpha^\beta
,
\quad
e^i{}_\alpha e_j{}^\alpha=\delta^i_j
.
\end{equation}
The position of the indices allows us to differentiate between the frame and its inverse without using different notation.

To progress further we need to measure the gravitational field strength.  It seems natural to differentiate the first expression from (\ref{1:3}) with respect to the coordinate $x^\gamma$, say, which gives
\begin{equation}
e^i{}_\alpha \partial_\gamma e_i{}^\beta
+
e_i{}^\beta \partial_\gamma e^i{}_\alpha
=0
.
\end{equation}
This motivates the introduction of the rank-3 tensor, antisymmetric in the first and third indices
\begin{equation}
\label{1:4}
K_{\alpha\beta\gamma}=g_{\gamma\gamma'}K_{\alpha\beta}{}^\gamma:=g_{\gamma\gamma'}e^i{}_\alpha \partial_\beta e_i{}^{\gamma'},
\end{equation}
which we take as our measure of field strength and call the \emph{contortion} tensor.

\section{Our approach}

So far we have introduced the frame $e^i{}_\alpha$ and linked this to the metric (\ref{1:2}).  We then identified a possible gravitational field strength tensor (\ref{1:4}).  The next logical step is to construct an admissible Lagrangian from which we can determine a set of field equations via a variational principle.  Here `admissible' means we take a very general Lagrangian and apply two physical assumptions to isolate physically meaningful situations.  These assumptions are discussed below.

As the reader will no doubt have noticed, we aim to model the gravitational field in a manner very different to general relativity.  Indeed, this is a good time to emphasise that we are \emph{not} considering a geometry with curvature.  If we assume we work in flat space then our metric (\ref{1:2}) is constant, and hence raising and lowering indices commutes with partial differentiation.  We need no longer differentiate between co- and contra-variant vectors and can work with Cartesian tensors, which reduces computational complexity significantly.  Stipulating that space-time is flat does \emph{not} remove gravity from our theory; it only serves to reduce the metric equation (\ref{1:2}) to a restrictive constraint on the frame components $e^i{}_\alpha$.

\section{Cautionary notes}

We stress that our treatment of general relativity via continuum mechanics is often referred to as the `teleparallel equivalent of general relativity', see~\cite{EinsteinCartan}. In this context, the free parameters of a general continuum-based theory are chosen to produce an equivalence with general relativity.

Our approach is different.  We do \emph{not} choose specific parameters to force such an equivalence; we retain all degrees of freedom and apply two physical assumptions which reduce our continuum model to general relativity.  It is the fact that two sensible physical assumptions alone produce general relativity that motivates this essay.

\section{Lagrangian and assumptions}

Define the Lagrangian $W$ with the rank-6 tensor $C_{\alpha\beta\gamma\lambda\mu\nu}$ as
\begin{equation}
\label{5:1}
W:=
\frac12
C_{\alpha\beta\gamma\lambda\mu\nu}
K_{\alpha\beta\gamma}
K_{\lambda\mu\nu}
\end{equation}
where $C$ takes all permutations over $(\alpha\beta\gamma\lambda\mu\nu)$, which has 4096 possible components.  Using symmetry arguments we will reduce this to something more manageable.  The symmetry properties of $K$ imply $C$ enjoys two symmetries, namely
\begin{equation}
\label{5:2}
C_{\alpha\beta\gamma\lambda\mu\nu}
=
-
C_{\gamma\beta\alpha\lambda\mu\nu}
,\quad
C_{\alpha\beta\gamma\lambda\mu\nu}
=
-
C_{\alpha\beta\gamma\nu\mu\lambda}
.
\end{equation}
Under (\ref{5:2}) the number of independent components of $C$ drops from 4096 to 576.  Irrespective of the symmetry properties of $K$, we also have
\begin{equation}
\label{5:3}
C_{\alpha\beta\gamma\lambda\mu\nu}
=
C_{\lambda\mu\nu\alpha\beta\gamma}
.
\end{equation}
Including (\ref{5:3}) reduces the number of independent components from 576 to 300.  We now state our two physical assumptions.
\begin{enumerate}
\item Isotropy: the property of a vacuum space-time whereby gravity acts equally in all directions.
\item Lorentz invariance: the property of a space-time whereby experimental results are independent of the orientation or the relative velocity of the laboratory through space.
\end{enumerate}
Both assumptions are well supported by current experimental evidence and are key to modern field theory.

Isotropy is invoked mathematically by stipulating that the tensor $C$ remains invariant under local changes of coordinates and, indeed, must be composed from tensors which obey the same restriction.  This means that a rank-6 tensor of our type can only be composed from Kronecker $\delta$'s.  A simple combinatorial exercise demonstrates there are 15 possible ways of choosing pairs from a set of 6 possibilities, which indicates that
\begin{equation}
\label{5:4}
W
=
\frac12\sum_{m=1}^{15}c_m\delta_{\textbf{m}(\alpha\beta\gamma\lambda\mu\nu)}
\end{equation}
where $c_m$ are undetermined parameters and $\textbf{m}(\alpha\beta\gamma\lambda\mu\nu)$ is one of the 15 possible combinations of indices.  To see the form of the isotropic tensors $\delta_{\textbf{m}(\alpha\beta\gamma\lambda\mu\nu)}$, consider the case when the indices retain their original order, indicated by $m=1$, say:
\begin{equation}
c_1\delta_{\textbf{1}(\alpha\beta\gamma\lambda\mu\nu)}
=
c_1\delta_{\alpha\beta}\delta_{\gamma\lambda}\delta_{\mu\nu}.
\end{equation}
We have not yet included the three symmetries (\ref{5:2}), (\ref{5:3}).  Doing so reduces the number of independent components to 3!  Explicitly we have
\begin{equation}
\label{5:5}
W
=
c_1
K_{\alpha\alpha\beta}K_{\beta\gamma\gamma}
+
c_2
K_{\alpha\beta\gamma}K_{\alpha\gamma\beta}
+
c_3
K_{\alpha\beta\gamma}K_{\alpha\beta\gamma}
\end{equation}
where $c_1$, $c_2$ and $c_3$ are the remaining free parameters.  We must now include the second physical assumption, Lorentz invariance. This manifests itself as the invariance of our Lagrangian under a certain transformation of the frame, namely
\begin{equation}
e^i{}_\alpha\mapsto \Lambda^i{}_j e^j{}_\alpha
\end{equation}
where $\Lambda^i{}_j(x^\kappa,t)$ is a member of the Lorentz group satisfying
\begin{equation}
\eta_{ij}\Lambda^i{}_m\Lambda^j{}_n=\eta_{mn}
.
\end{equation}
Our metric (\ref{1:2}) is invariant under such a transformation.  The required invariance of the Lagrangian (\ref{5:5}) produces a constraint on the undetermined parameters $c_1$, $c_2$ and $c_3$, specifically that
\begin{equation}
c_1=-c_2
,
\qquad
c_3=0
.
\end{equation}
Therefore we have reduced the number of independent components of $C$ from 4096 to 1, and only 1, free parameter!

\section{The final result}

Our one-parameter Lagrangian can now be written
\begin{equation}
W=c_1
\left(
K_{\alpha\alpha\beta}K_{\beta\gamma\gamma}
-
K_{\alpha\beta\gamma}K_{\alpha\gamma\beta}
\right)
.
\end{equation}
This is a significant achievement.  We have produced a simple Lagrangian from a very general case using two physical assumptions, all without introducing curvature.  Furthermore, one can show that our Lagrangian is, up to a surface term, equivalent to the Einstein-Hilbert action of general relativity.

At this point we could vary our Lagrangian to determine a set of field equations.  These will be equivalent to the Einstein field equations, although expressed in terms of the tetrad instead of the metric.

As a final thought, it is possible to introduce different physical assumptions into our model by replacing isotropy with another symmetry. This asymmetry is at a very fundamental level of the theory and is completely independent of the symmetries of the metric. One can formulate an anisotropic model ($C_{\alpha\beta\gamma\lambda\mu\nu}$ anisotropic) where the space-time ($g_{\alpha\beta}$) would remain isotropic for instance.  Our model allows for the straightforward introduction of such non-standard symmetries.

Our essay demonstrates a very quick method of arriving at general relativity using the machinery of continuum mechanics.  It also provides a neat method for investigating symmetries of the gravitational field and, hopefully, the reader has not at any point felt as though they were ``swimming underwater, blindfolded, and with [their] hands tied behind [their] back[s]"!

\subsection*{Acknowledgements}
We thank Dmitri Vassiliev and Caitlin Dempsey for useful discussions.

\end{document}